\documentclass[10pt,letterpaper]{article}
\usepackage{opex3}
\usepackage{epsfig}
\usepackage{amsmath}
\usepackage{units}

\newcommand{\D}{\mathrm{d}}
\newcommand{\E}{\mathrm{e}}
\newcommand{\bra}[1]{\left< #1 \right|}
\newcommand{\ket}[1]{\left| #1 \right>}

\begin{document}

\title{Amplitude and phase modulation of time-energy entangled two-photon states}

\author{F.~Z\"ah$^{1}$, M.~Halder$^{2}$, and T.~Feurer$^{1}$}

\address{$^{1}$University of Bern, Institute of Applied Physics,
Sidlerstr. 5, 3012 Bern, Switzerland $^{2}$Universit\'e de Gen$\grave{e}$ve,
GAP-Optique, Rue de l'\'Ecole-de-M\'edecine 20, 1211 Gen$\grave{e}$ve,
Switzerland}

\email{florian.zaeh@iap.unibe.ch}



\begin{abstract}
We experimentally demonstrate amplitude and phase modulation of a time-energy
entangled two-photon wave function. The entangled photons are produced by
spontaneous parametric down-conversion, spectrally dispersed in an prism
compressor, modulated in amplitude and/or phase, and detected in coincidence by
sum-frequency generation. First, we present a Fourier optical analysis of the
optical setup yielding an analytic expression for the resulting field
distribution at the exit plane of the shaping apparatus. We then introduce
amplitude and/or phase shaping and present results which can only be obtained
through a combination of the two. Specifically, we use a shaper-based
interferometer to measure the two-photon interference of an almost
bandwidth-limited two-photon wave function.
\end{abstract}



\section{Introduction}

Entangled photon states are ideal subjects to study nonlocal interactions and
applications in quantum communication or quantum information processing
\cite{Schrodinger,Popescu,Gisin,Zeilinger-Springer}. A convenient way to
produce such states is through spontaneous parametric down-conversion (SPDC).
An intense pump beam creates a second order nonlinear response strong enough to
facilitate the annihilation of a pump photon and the creation of a pair of
down-converted photons. In a type-I process the photon pairs consist of a
signal and an idler photon and show entanglement with respect to space-wave
vector and time-energy \cite{Weinberg}. In the past these photon pairs have
been used for example to demonstrate the violation of Bell's inequality
\cite{Mandel,Alley}, to absolutely calibrate single photon counters
\cite{Sergienko,Petroff}, to demonstrate the appearance of fourth order
interference in absence of second order interference \cite{Ghosh,Kwiat}, to
explore two-photon imaging \cite{Shih,Pittman,Yamamoto,Dowling}, or to
investigate fundamental properties of entangled photon pairs
\cite{Weinberg,HOM,Franson,Lu}. Recently, it was demonstrated that the
visibility of a fourth order interferogram of the idler beam can be affected by
a spectral bandpass filter in the signal beam \cite{Ital_short,Ital_long}, and
that the quantum state of a photon pair can be phase-modulated in the same way
as coherent classical light pulses are tailored \cite{Sil05_1,Sil05_2}. In
order to affirm the effect of phase modulation on the two-photon wave function,
coincidences were detected through sum-frequency generation. This detection
scheme has a rather low efficiency but became viable because the short
coherence time of the photon pairs allowed for a high flux while remaining in
the single photon limit \cite{Sil05_1,Sil05_2}.

Here, we use a very similar setup as the one presented by Silberberg and
coworkers \cite{Sil05_1} but extend phase-only shaping of the spectral
components to phase and/or amplitude shaping. First, we present a
Fourier-optical analysis of the optical setup and give an analytic expression
for the field distribution at the exit plane of the shaping apparatus which
also coincides with the position of the coincidence detection. This result is
subsequently used to simulate all experimental results. We then proceed by
demonstrating the increased versatility of the setup for example by measuring
the second order interferogram of a bandwidth-limited two-photon wave function
through a shaper-based interferometer.

\section{Experimental realization}

The experimental setup was similar to that reported by the Silberberg group
\cite{Sil05_1}. The entangled photon pairs were created through SPDC of a
continuous wave \unit[532]{nm} single-mode pump laser in a temperature
stabilized, periodically poled KTiOPO$_4$ (PPKTP) crystal. Both photons had the
same polarization and the entanglement was with respect to time-energy. The
spectral bandwidth of the pump laser was approximately \unit[5]{MHz} and the
maximum pump power \unit[5]{W}. Given a poling periodicity of $G
=$~\unit[9]{$\mu$m}, phase matching allowed for generating an approximately
\unit[50]{nm} broad spectrum centered at twice the pump wavelength. The exact
shape and bandwidth of the spectrum was dominated by the crystal temperature.
Here, the temperature was set to \unit[29.5]{$^\circ$C} maximizing the spectral
bandwidth as well as the conversion efficiency, which was on the order of
$10^{-7}$. The focusing lens for the pump laser was selected according to the
optimum focusing condition \cite{boyd68} and had a focal length of
\unit[150]{mm}. The emerging photon pairs were imaged to an intermediate plane
and from there to a second PPKTP crystal. Both imaging sections included a
two-prism combination, first, to spectrally disperse the down-converted
spectrum at the intermediate plane and, second, to compensate for any second
order dispersion in the setup. A computer-controlled spatial light modulator
(JenOptik SLM 640-d) was placed at the intermediate plane and modulated the
amplitude and the phase of selected spectral components. The modulated spectrum
was recombined in the second crystal where it generated sum-frequency photons
with a maximum efficiency on the order of $10^{-9}$. A spectral filter (4mm
BG18) suppressed the remaining photon pairs and the sum-frequency photons were
collected by a multi-mode fiber connected to a single-photon counter
(PerkinElmer SPCM-AQR-15). Its efficiency at \unit[532]{nm} is about 33 times
higher than at \unit[1064]{nm}. Detecting the sum-frequency photons was
essentially similar to a coincidence detection scheme with an extremely high
temporal resolution \cite{Sil05_1}. Almost all measurements used an integration
time between \unit[5]{s} and \unit[100]{s} and the dark count rate was
approximately \unit[60]{cps}. In order to verify that the detector was
measuring sum-frequency photons at \unit[532]{nm} rather than residual photons
around \unit[1064]{nm} we increased the temperature of the second crystal by a
few degrees, thus detuning the acceptance function of the sum-frequency
crystal, and observed that the count rate dropped down to the dark count rate.

Because the optical setup is quite different from standard 4f-type geometries
\cite{Wei00} we first present a detailed Fourier optical analysis. We use the
paraxial approximation and assume that the optical axis is parallel to the $z$
axis. It is sufficient to treat the problem in one dimension, $x$, as the other
dimension, $y$, remains unaffected by the prism pairs. We use the transfer
functions for free-space propagation $T_z(k_x,\omega) = \exp\left[ -i k(\omega)
z + i \frac{k_x^2}{2k(\omega)} z \right]$, for an ideal lens $T_L(x,\omega) =
\exp\left[ \frac{i k(\omega)}{2 f} x^2 \right]$, and for a prism in minimum
deviation geometry $T_P(x,\omega) = \exp\left[ i \gamma (\omega-\omega_c) x
\right]$, with the center frequency of the optical wavepacket $\omega_c$, its
center wavelength $\lambda_c$, and $\gamma \approx -\frac{2\lambda_c}{c} \left.
\frac{\D n}{\D\lambda} \right|_{\lambda_c}$.

\begin{figure}[htb]
\centering \epsfig{file=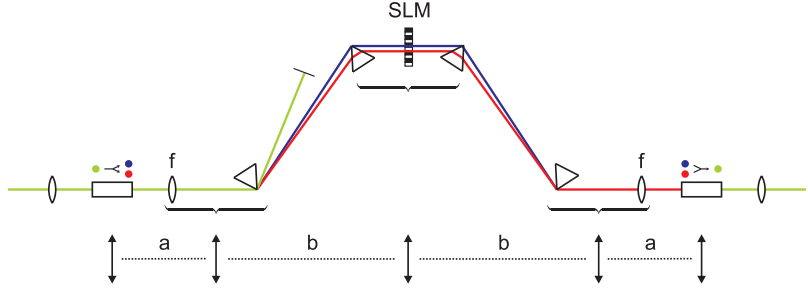,width=0.8\columnwidth}

\caption{\label{fig:setup_images} Optical setup and schematic for Fourier
optical analysis.}
\end{figure}

Figure~\ref{fig:setup_images} shows a schematic of the setup. Both imaging
lenses have a focal length of $f$ and the distances crystal--lens and lens--SLM
are $a$ and $b$, respectively, yielding a magnification of $m=b/a$. Without
loss of generality, we assume that the first prism follows immediately after
the imaging lens and the second prism is located right before the intermediate
SLM plane. Then, we find for a classical light field in the last image plane,
i.e. at the position of the up-conversion crystal

\begin{equation}\label{finalfield}
 \widetilde{E}_1^+(x,\omega) \propto \widetilde{E}_0^+(x,\omega) \; M\left(
 -m x - \frac{f (m+1)}{k_c} \gamma (\omega-\omega_c) \right) \; \exp\left[
 -i \frac{k_c m}{f} x^2 + i \frac{f (m+1)}{k_c} \gamma^2 (\omega-\omega_c)^2
 \right],
\end{equation}

with the electric field at the origin $\widetilde{E}_0^+(x,\omega)$ and the
wave vector $k_c \doteq k(\omega_c)$. Note that the size and position of the
resulting field are independent of the phase and amplitude modulation applied,
however, the spectral modulation is a function of space and a curved phase is
introduced. The first term of the exponential function specifies a quadratic
spatial phase which is related to the imaging geometry. The second part is
quadratic in $(\omega-\omega_c)$ and reflects the group velocity dispersion of
the four prism arrangement. Keeping in mind that the distance between the two
prisms is $b = f (m+1)$ and inserting the explicit expression for $\gamma$, it
is easy to show that the quadratic phase corresponds to the second order Taylor
coefficient of a regular prism compressor \cite{Diels}. In our setup the
distance between the two prisms is adjusted such that all positive dispersion
present in the optical setup is compensated for and the net dispersion is zero.
The argument of the modulator's transfer function $M(x)$ depends both on $x$
and $\omega$, which is a consequence of space-time coupling. The
frequency-to-space mapping is given by $x = f (m+1) \gamma (\omega-\omega_c) /
k_c$. Assuming that the incoming field may be decomposed into a spatial and a
spectral part $\widetilde{E}_0(x,\omega) = F(x) \; \widetilde{E}_0(\omega)$
and, further, that only space averaged fields are measured yields

\begin{equation}\label{xaverage}
 \widetilde{E}_1(\omega) = \widetilde{E}_0(\omega) \; \mathcal{M}(\omega),
\end{equation}

with

\begin{equation}\label{Mw}
 \mathcal{M}(\omega) \propto \int \D x \; F(x) \; M\left(
 -m x - \frac{f (m+1)}{k_c} \gamma (\omega-\omega_c) \right) \; \exp\left(
 -i \frac{k_c m}{f} x^2 \right).
\end{equation}

\section{Quantum optical description of the measurement}

In the following section we briefly review the wave function which replaces the
classical field in Eq.~\ref{xaverage} and which is modulated by a specific
modulator transfer function $\mathcal{M}(\omega)$. While the pump field is
treated classically, the signal and the idler fields are quantized. Signal and
idler photons have the same polarization and, thus, experience the same index
of refraction. The two-photon wave function generated in SPDC has been derived
in reference \cite{Rubin} with first order perturbation theory. In all
experiments reported here, the frequency resolution at the intermediate plane
rather than the bandwidth of the pump laser limits the coherences observed,
that is, the pump field can be approximated by a monochromatic field
$\widetilde{E}_p(\omega) = \widetilde{E}_p \; \delta(\omega-\omega_{pc})$ with
a frequency $\omega_{pc}$ and the two-photon wave function is

\begin{equation}\label{Psi2}
 \ket{\Psi} = \ket{0} + \int \D\omega_s \; \xi(\omega_s) \;
 \widehat{a}_s^\dagger(\omega_s) \; \widehat{a}_i^\dagger(\omega_{pc}-\omega_s)
 \; \ket{0},
\end{equation}

with

\begin{equation}\label{chi2}
 \xi(\omega_s) \doteq \alpha \widetilde{E}_p \; \mathrm{sinc}\left(
 \frac{\Delta k L}{2} \right) \; \E^{-i \Delta k \; L/2}.
\end{equation}

All constants and slowly varying dependencies are combined in $\alpha$, $\Delta
k \doteq k_p(\omega_p) - k_s(\omega_s) - k_i(\omega_i) - 2\pi/G$ is the phase
mismatch, and $L$ the length and $G$ the periodicity of the periodically poled
crystal. The sum-frequency signal measured after the second crystal is
proportional to the second order coherence function assuming that for a
perfectly aligned setup there is no delay between the signal and the idler
photon. That is, the coincidence rate is given by

\begin{equation} \label{coincidence}
 G^{(2)}(0,0) = \left| \bra{0} \widehat{E}_s^+(0) \; \widehat{E}_i^+(0)
 \ket{\Psi} \right|^2 = \left| \int \D\omega_s \; M_s(\omega_s) \;
 M_i(\omega_{pc}-\omega_s) \; \xi(\omega_s) \right|^2.
\end{equation}

If the pulse shaper is used then $M_s(\omega_s) \; M_i(\omega_{pc}-\omega_s) =
\mathcal{M}(\omega_s)$.

\section{Experimental results}

All experiments presented in the following were simulated by numerically
solving Eq.~\ref{coincidence} with the appropriate modulator transfer function
Eq.~\ref{Mw}. The simulations take into account the pixelated nature of the SLM
used and assume an uncertainty of the temperature of $\pm 1$~K and of the beam
waist of \unit[$\pm 30$]{\%}.

\begin{figure}[htb]
\centering \epsfig{file=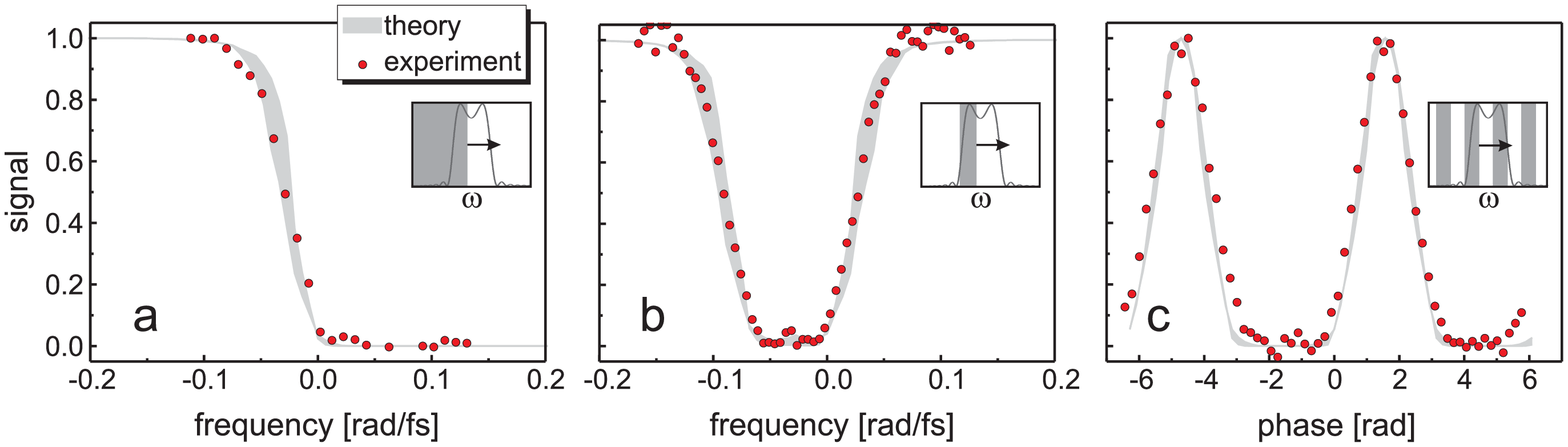,width=\columnwidth}

\caption{\label{fig:amplitude} Coincidence signal as a function of the position
of (a) a spectral edge, (b) a spectral slice, and (c) of an amplitude grating.
The insets show the shape of the different amplitude filters which are scanned
across the spectrum.}
\end{figure}

First, we present three different pure amplitude-only modulation experiments
which imply $\arg[M_\mathrm{s,i}(\omega)] = 0$. Figure~\ref{fig:amplitude}(a)
shows the signal versus the position of a spectral edge filter. By moving the
edge filter across the spectrum more and more frequencies are blocked and the
signal drops to zero. Note that zero signal is reached exactly when one half of
the spectrum is blocked; in other words, removing all idler photons from all
photon pairs is sufficient to destroy all coincidences measured at the second
crystal. When only a spectral slice is blocked and scanned across the spectrum,
the signal drops to a minimal value, exhibits a small peak around the center
frequency, and then increases back to the initial value, as shown in
Fig.~\ref{fig:amplitude}(b). The signal is minimal when either most of the
idler or most of the signal photons are blocked. The height and shape of the
central peak depend on the width of the spectral block compared to the
spectrum. Here, the spectral block is almost half as wide as the spectrum and,
consequently, the peak is barely visible. Figure~\ref{fig:amplitude}(c) shows
the signal as a function of the position of an amplitude grating. The signal
exhibits the same periodicity as the grating, is zero when the grating is
asymmetric and maximal when the grating is symmetric with respect to the center
frequency. In the asymmetric case the idler photons of one half of all photon
pairs and the signal photons of the other half of all photon pairs are blocked
causing the overall coincidence signal to disappear. Although the overall
number of photons has been reduced by only one half, the signal is zero because
not a single photon pair is left intact. In the symmetric case the amplitude
grating blocks all idler and all signal photons of the same photon pairs, but
passes all other photon pairs unaffected.

\begin{figure}[htb]
\centering \epsfig{file=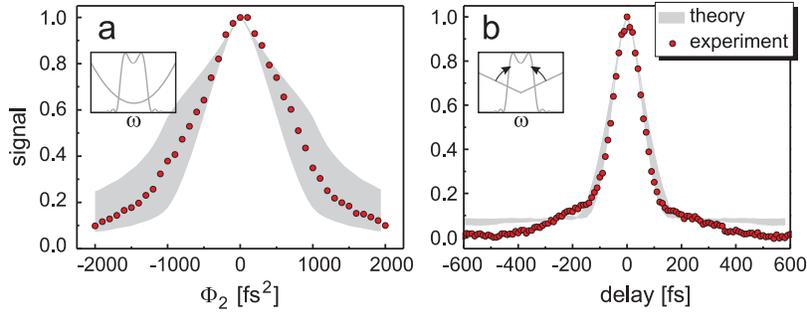,width=0.8\columnwidth}

\caption{\label{fig:phase} Coincidence signal as a function of (a) the
magnitude of the quadratic spectral phase and (b) the slope of a V-shaped
linear phase. The insets show the shape of the two types of phase modulations.}
\end{figure}

For pure phase modulations the amplitude of the modulator transfer function is
$\left|M_\mathrm{s,i}(\omega)\right| = 1$. Various phase-only modulation
examples have already been published by the Silberberg group
\cite{Sil05_1,Sil05_2,Sil_07} and two examples are shown in
Fig.~\ref{fig:phase}. When the phase modulation is quadratic in frequency, i.e.
$M_\mathrm{s,i}(\omega) = \exp[i \Phi_2/2 (\omega-\omega_{pc}/2)^2]$, the
two-photon wave function is smeared out in time. In such a broadened two-photon
wave function the chances for a coincidence are reduced resulting in a
decreasing signal with increasing $|\Phi_2|$, as seen in
Fig.~\ref{fig:phase}(a). The result is similar for the classical light field of
a coherent short pulse. Observing the maximum signal at $\Phi_2 = 0$ confirms
that the four-prism arrangement has been aligned such that it compensates for
all positive quadratic dispersion. Next, we apply a V-shaped phase modulation.
Such a phase modulation was already used in reference \cite{Sil05_1} and it
shifts the idler photon with respect to the signal photon and samples the
amplitude of the Fourier transform of $\xi(\omega)$. From the result presented
in Fig.~\ref{fig:phase}(a) we can deduce a coherence time of the two-photon
wave function of approximately \unit[150]{fs}. In both experiments the
theoretical predictions agree well with the experimental results.

\begin{figure}[htb]
\centering \epsfig{file=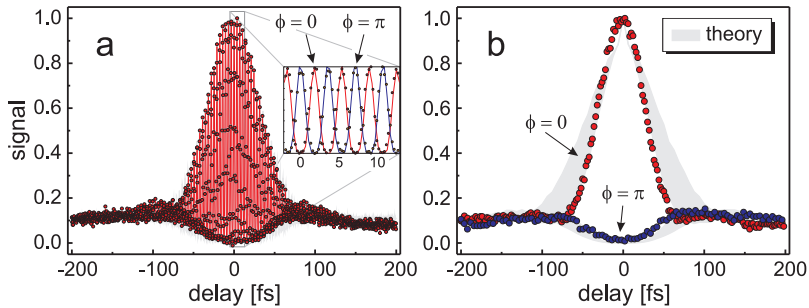,width=0.8\columnwidth}

\caption{\label{fig:autocorr} Two-photon interference with (a) $\gamma=1$ and
$\phi=0,\pi$, and (b) $\gamma=0$ and $\phi=0,\pi$. The inset in (a) shows a
small section of the two-photon interference for $\phi=0$ and $\pi$,
respectively.}
\end{figure}

The last experiment requires simultaneous amplitude and phase shaping and our
intention is to demonstrate a pulse shaper based unbalanced interferometer in
front of the coincidence detector and to measure the two-photon interference.
For a standard unbalanced interferometer, i.e. a Michelson or a Mach-Zehnder
interferometer, the transfer function is

\begin{equation}\label{inter1}
 M_\mathrm{s,i}(\omega) = r t \left( 1 + \E^{-i \omega \tau} \right),
\end{equation}

with the reflectivity $r$ and the transmission $t$ of the beamsplitter and the
time delay $\tau$. A pulse shaper allows mimicking a much more flexible
transfer function, such as

\begin{equation}\label{inter2}
 M_\mathrm{s,i}(\omega) = \frac{1}{2} \; \left[ 1 + \exp\left( -i \omega
 \tau + i (1-\gamma) \frac{\omega_{pc}}{2} \tau - i \phi \right) \right].
\end{equation}

We see that for $\gamma=1$ and $\phi=0$ eq.~\ref{inter2} resembles
eq.~\ref{inter1} and an unbalanced Michelson interferometer may be simulated.
The measured coincidence rate versus time delay, i.e. the two-photon
interference, for $\gamma=1$ and $\phi=0$ is shown Fig.~\ref{fig:autocorr}(a).
The signal oscillates with a periodicity that is determined by the frequency
$\omega_{pc}/2$. The inset indicates that by selecting $\phi=0$ or $\phi=\pi$
allows to switch between the two output ports of the simulated interferometer;
the two signals are exactly half a period out of phase. In order to measure
these two signals with a real Michelson interferometer would require to move
the coincidence detection apparatus from one exit port of the beam splitter to
the other. If we select $\gamma=0$, only the slowly varying amplitudes of the
signal and idler photons are delayed in time leaving their carrier frequencies
unaffected. The results change quite dramatically, i.e. the oscillations
completely disappear, as seen in Fig.~\ref{fig:autocorr}(b). The results in
Fig.~\ref{fig:autocorr}(b) are especially interesting, because the two curves
can readily be used to extract the fringe visibility as a function of time
delay. From both measurements we can extract the coherence properties of the
two-photon wave function, which must be close to bandwidth-limited because all
second order dispersion has been compensated for.

\section{Conclusion}

We have presented a Fourier-optical analysis of the shaping setup first
introduced to shaping of two-photon wave functions by the Silberberg group. By
extending phase-only modulation to phase and amplitude modulation we have shown
that a much larger variety of transfer functions can be realized. Specifically,
we demonstrated a shaper-assisted unbalanced Michelson interferometer without
any moving parts and measured the two-photon interference with it. With all
second order dispersion removed by the four prism arrangement the coherence
properties should be close to bandwidth limited. Lastly, we have demonstrated a
measurement that cannot be obtained with a mechanical unbalanced Michelson
interferometer and which yields and 'oscillation-free' two-photon interference
from which the fringe visibility can be readily derived.

\section*{Acknowledgments}

We thank A.~Pe'er, B.~Dayan, and Y.~Silberberg for many discussions. This work
was supported by NCCR Quantum Photonics, research instrument of the Swiss
National Science Foundation.

\begin{thebibliography}{99}


\bibitem{Schrodinger} E.~Schr\"{o}dinger, "Die gegenw\"artige Situation
    in der Quantenmechanik," Naturwissenschaften \textbf{23}, 807--812,
    823--828, 844--849 (1935)

\bibitem{Popescu} S.~Popescu, and D.~Rohrlich, "The joy of
    entanglement," in \emph{Introduction to quantum computation
    and information}, H.-K.~Lo, S.~Popescu, T.~Spiller, eds. (World
    Scientific, 1998), pp. 29--48

\bibitem{Gisin} N.~Gisin, and R.~Thew, "Quantum communication,"
    Nature Phot. {1}, 165--171 (2007)

\bibitem{Zeilinger-Springer} D.~Bouwmeester, A.~Ekert, and A.~Zeilinger
    \emph{The Physics of Quantum Information} (Springer, 2000)

\bibitem{Weinberg} D.C.~Burnham, and D.L.~Weinberg, "Observation
    of simultaneity in parametric production of optical photon pairs,"
    Phys. Rev. Lett. \textbf{25}, 84--87 (1970)


\bibitem{Mandel} Z.Y.~Ou, and L.~Mandel, "Violation of Bell's inequality
    and classical probability in a two-photon correlation experiment",
    Phys. Rev. Lett. \textbf{61}, 50--53 (1988)

\bibitem{Alley} Y.H.~Shih, and C.O.~Alley, "New type of
    Einstein-Podolsky-Rosen-Bohm experiment using pairs of light quanta
    produced by optical parametric down conversion," Phys. Rev. Lett
    \textbf{61}, 2921--2924 (1988)


\bibitem{Sergienko} A.A.~Malygin, A.N.~Penin, A.V.~Sergienko, "Absolute
    calibration of the sensitivity of photodetectors using a biphotonic
    field", Sov. Phys. JETP Lett. \textbf{33}, 477--480 (1981)

\bibitem{Petroff} P.G.~Kwiat, A.M.~Steinberg, R.Y.~Chiao,
    P.H.~Eberhard, and M.D.~Petroff, "High-efficiency single-photon
    detectors," Phys. Rev. A \textbf{48}, 867--870 (1993)

\bibitem{Ghosh} R.~Ghosh, and L.~Mandel, "Observation of nonclassical
    effects in the interference of two photons," Phys. Rev. Lett.
    \textbf{59}, 1903--1905 (1987)

\bibitem{Kwiat} P.G.~Kwiat, W.A.~Vareka, C.K.~Hong, H.~Nathel, and
    R.Y.~Chiao, "Correlated two-photon interference in a dual-beam
    Michelson interferometer," Phys. Rev. A \textbf{41}, 2910--2913 (1990)


\bibitem{Shih} D.V.~Strekalov, A.V.~Sergienko, D.N.~Klyshko, and
    Y.H.~Shih, "Observation of Two-Photon "Ghost" Interference and
    Diffraction," Phys. Rev. Lett. \textbf{74}, 3600--3603 (1995)

\bibitem{Pittman} T.B.~Pittman, Y.H.~Shih, D.V.~Strekalov, and
    A.V.~Sergienko, "Optical imaging by means of two-photon quantum
    entanglement," Phys. Rev. A \textbf{52}, 3429--3432 (1995)


\bibitem{Yamamoto} J.~Jacobson, G.~Bj\"ork, I.~Chuang, and Y.~Yamamoto,
    "Photonic de Broglie waves," Phys. Rev. Lett. \textbf{74}, 4835--4838
    (1995)

\bibitem{Dowling} A.N.~Boto, P.~Kok, D.S.~Abrams, S.L.~Braunstein,
    C.P.~Williams, and J.P.~Dowling, "Quantum interferometric optical
    lithography: exploiting entanglement to beat the diffraction limit,"
    Phys. Rev. Lett. \textbf{85}, 2733--2736 (2000)


\bibitem{HOM} C.K.~Hong, Z.Y.~Ou, and L.~Mandel "Measurement of
    Subpicosecond Time Intervals between Two Photons by Interference,"
    Phys. Rev. Lett. \textbf{59}, 2044--2046 (1987)

\bibitem{Franson} J.~D.~Franson, "Bell inequality for position and time,"
    Phys. Rev. Lett. \textbf{62}, 2205--2208 (1989)

\bibitem{Lu} Z.Y.~Ou, and Y.J.~Lu, "Cavity enhanced spontaneous
    parametric down-conversion for the prolongation of correlation time
    between conjugate photons," Phys. Rev. Lett. \textbf{83}, 2556--2559
    (1999)

\bibitem{Ital_short} M.~Bellini, F.~Marin, S.~Viciani, A.~Zavatta, and
    F.T.~Arecchi, "Nonlocal Pulse Shaping with Entangled Photon Pairs," Phys.
    Rev. Lett. \textbf{90}, 043602 (2003)

\bibitem{Ital_long} S.~Viciani, A.~Zavatta, and M.~Bellini, "Nonlocal
    modulations of the temporal and spectral profiles of an entangled photon
    pair," Phys. Rev. A \textbf{69}, 053801 (2004)

\bibitem{Sil05_1} B.~Dayan, A.~Pe'er, A.A.~Friesem, and Y.~Silberberg,
    "Nonlinear Interactions with an Ultrahigh Flux of Broadband Entangled
    Photons," Phys. Rev. Lett. \textbf{94}, 043602 (2005)

\bibitem{Sil05_2} B.~Dayan, A.~Pe'er, A.A.~Friesem, and Y.~Silberberg,
    "Temporal Shaping of Entangled Photons," Phys. Rev. Lett. \textbf{94},
    073601 (2005)

\bibitem{boyd68} G.D.~Boyd, and D.A.~Kleinman, "Parametric Interaction of
    Focused Gaussian Light Beams," J. Appl. Phys. \textbf{39}, 3597--3639
    (1968)

\bibitem{Wei00} A.M.~Weiner, "Femtosecond pulse shaping using spatial light
    modulators," Rev. Sci. Instrum. \textbf{71}, 1929--1960 (2000)

\bibitem{Diels} J.C.~Diels, and W.~Rudolph, \emph{Ultrashort Laser Pulse
    Phenomena} (Academic Press, 1996)

\bibitem{Rubin} T.E.~Keller, and M.H.~Rubin, "Theory of two-photon
    entanglement for spontaneous parametric down-conversion driven by a narrow
    pump pulse," Phys. Rev. A \textbf{56}, 1534--1541 (1997)

\bibitem{Sil_07} B.~Dayan, Y.~Bromberg, I.~Afek, and Y.~Silberberg, "Spectral
    polarization and spectral phase control of time-energy entangled photons,"
    Phys. Rev. A \textbf{75}, 043804 (2007)
\end{thebibliography}
\end{document}